# Enabling New ALMA Science with Improved Support for Time-Domain Observations

The ALMA Time-domain Special Interest Group[1]

**Abstract.** While the Atacama Large Millimeter/submillimeter Array (ALMA) is a uniquely powerful telescope, its impact in certain fields of astrophysics has been limited by observatory policies rather than the telescope's innate technical capabilities. In particular, several observatory policies present challenges for observations of variable, mobile, and/or transient sources — collectively referred to here as "time-domain" observations. In this whitepaper we identify some of these policies, describe the scientific applications they impair, and suggest changes that would increase ALMA's science impact in Cycle 6 and beyond.

## 1 Introduction

The Atacama Large Millimeter/submillimeter Array (ALMA; Brown et al. 2004) is recognized as a uniquely powerful telescope, and bibliographical studies to date indicate that ALMA observations are published at a high rate (Stoehr et al. 2015). It is also true, however, that ALMA is a young observatory that has not yet achieved its full potential — as demonstrated by the ongoing ALMA Development Program, which funds the development of new capabilities that will increase ALMA's ability to produce cutting-edge scientific results.

This whitepaper is concerned with ALMA's "time-domain" capabilities, where we use this term to refer to observations that involve astronomical sources that are variable, mobile, and/or transient. As we show below, time-domain observations at mm/sub-mm wavelengths are important in a wide variety astrophysical topics, ranging from solar system science to the study of distant cosmic explosions.

Below, we identify three areas in which ALMA's time-domain capabilities are limited relative to other major observatories:

- Time-critical observations, including observations coordinated with other telescopes
- Rapid-turnaround target-of-opportunity observations
- Long, continuous observations

We present the science drivers that motivate improved capabilities and provide specific recommendations for actions to be taken by the Joint ALMA Observatory and/or the ALMA Partnership, hereafter referred to simply as "ALMA." In all cases, the barriers that we identify are not due to limitations in the telescope's innate capabilities, but rather

---

[1]A full list of ATSIG members may be found at the end of this whitepaper. Corresponding author: P. K. G. Williams, pwilliams@cfa.harvard.edu.

ALMA policy choices. ALMA's time-domain capabilities can therefore be improved substantially, unlocking its ability to deliver forefront science in a variety of fields, at costs that we anticipate to be much lower than those associated with the ALMA Development Program.

## 2 Time-Critical Observations

We define "time-critical observations" as those that must be executed in a specific time frame that is known well in advance of the time of observation. (This definition is somewhat different than that implied in the ALMA Cycle 4 Proposer's Guide [C4PG].) While these can include observations of rare astrophysical events or observing geometries, we believe that the most common motivation for time-critical observations is the desire to coordinate with another observatory.

The C4PG states that "Time-critical observations requiring a time window smaller than 14 days will not be guaranteed, but may be attempted on a best-effort basis … observations with strict timing constraints but many possible time windows may be feasible." Given this guidance, proposers can have no reasonable expectation that a TAC will approve a project whose science case depends on obtaining a time-critical observation unless it happens to have extremely loose timing constraints.

### 2.1 Science Drivers

Coordinated multi-wavelength observations are important tools in a variety of astrophysical contexts, ranging from studies of young stars (e.g., Cohen & Schwartz 1976) to microquasars (e.g., Mirabel & Rodríguez 1999) to active galactic nuclei (AGN; e.g., Markoff et al. 2008). Particularly relevant at mm wavelengths are studies of Sgr A*, the fluctuating source of radiation associated with the black hole at the center of the Milky Way. For instance, in 2016 July, *Chandra X-ray Observatory* and *Spitzer Space Telescope* performed two coordinated, 24-hour observations of Sgr A*, with a goal of understanding the physical origin of the source's flares. Simultaneous ALMA observations would reveal whether the mm emission originates in the same plasma responsible for the infrared and X-ray emission. While a DDT proposal allowed for simultaneous ALMA observations for 25% of each visit, *such an observation could not have been approved through the regular proposal process*. The Sgr A* observing project anticipates ~150 hr of simultaneous *Chandra/Spitzer* observations in the next two years, demonstrating the high priority that the community assigns to this work.

Magnetically active T Tauri stars that host protoplanetary disks are known to undergo large and rapid changes in their UV to X-ray emission as a result of flares. Flares can have rise times of less than an hour and can reach peak X-ray luminosities up to two orders of magnitude larger than quiescent values, with decay timescales of several hours to a day or more. The energetic radiation of T Tauri stars is a dominant source of ionization for protoplanetary disks and drives certain aspects of disk surface chemistry to which ALMA can be sensitive. Observations of time-dependent changes in chemical diagnostics can provide vital clues concerning this link and insights into the way disk chemistry depends on host star energetic radiation. Stellar flares are stochastic events and can only be caught serendipitously. While giants flares are rare, moderate fares are quite frequently observed in X-ray observations. Coordinated observations between ALMA and other observatories are required to study any resulting changes to the disk. The schedules of observatories such as *Chandra* and *XMM-Newton* must be fixed weeks in advance, so that coordinated ALMA observations require time-critical scheduling.

## 2.2 Recommended Policy Changes

We recommend that ALMA support all time-critical observations, regardless of the size of the proposed time window. We recognize that this support would make the observatory schedule more complex to determine and could impact overall observing efficiency. We therefore recommend that ALMA follow the lead of other observatories and explicitly allow proposers to request time-critical observations, but cap the total number of such requests that will be granted during each cycle at a level that facilitates an acceptable average observing efficiency. We recommend that the support for such proposals include ones whose precise time constraints cannot be known at the time of proposal preparation, if their timing can be fixed well in advance of the intended observation. This category would include proposals requesting coordinated observations with other observatories, which should be explicitly supported in the Proposer's Guide.

In addition, we urge ALMA to work to establish joint proposal agreements with other major astronomical observatories. As is broadly recognized, such agreements increase the participating observatories' scientific impact by encouraging proposals for coordinated multi-observatory projects, which otherwise face additional risks due their need for approval by two separate TACs. We further encourage ALMA to work to ensure that ALMA's proposal deadlines and observing seasons are well-aligned with the schedules of other major observatories.

# 3 Rapid-Turnaround Target-of-Opportunity Observations

Target-of-opportunity (ToO) observations are distinct from "time-critical" observations, as defined above, because the time of the observations cannot be determined well in advance. Their scientific return often depends on the requested observations being executed as soon as possible after a ToO trigger is issued. The C4PG states that "the Observatory will attempt to observe ToO proposals during the 48 hours following their triggering provided the appropriate scheduling conditions … are met. However, critical activities of the Observatory such as engineering and activities associated with the optimization and further development of the Array will not be interrupted …".

## 3.1 Science Drivers

A primary science driver for rapid-response ToO observations is the study of cataclysmic explosions such as $\gamma$-ray bursts (GRBs), supernovae (SNe), tidal disruption events (TDEs), and electromagnetic counterparts to gravitational wave detections. Below we emphasize that the scientific return of such observations often demands not only rapid turnaround on ToO scheduling, but also rapid analysis of the resulting data to allow planning of follow-up observations with both ALMA and other observatories.

Perhaps the most stringent requirements on response time come from GRB observations. Rapid response ($\Delta t < 7$ d) in the millimeter bands can probe reverse shock emission, which diagnoses the jet launching and collimation mechanism by constraining the ejecta magnetization, the initial Lorentz factor, and the composition of the ejecta (baryon or magnetically dominated; Meszaros & Rees 1993; Sari & Piran 1999). Millimeter-band observations at $\Delta t \leq 14$ d further probe the peak of the forward shock SED with larger separation between models than any other band, providing a unique measure of the density profile of the circumburst medium. Additionally, several expected physical effects at these time scales (such as energy injection reverse shocks) can only be studied with ALMA (Laskar et al. 2015). The mm band does not suffer from interstellar scintillation, thus clarifying the interpretation of cm-band observations, which may be crippled in the absence of higher frequency observations. Finally, millimeter-band spectroscopy of GRB afterglows may yield molecular absorption features from the host ISM, allowing for a unique probe of molecular gas in high-redshift galaxies (Inoue et al. 2007).

For core-collapse SNe, early observations can probe the characteristics of the seemingly prevalent, but poorly understood, phenomenon of late-stage progenitor mass loss (Smith 2014). Rapid-response observations of thermonuclear (Type Ia) SNe diagnose the progenitors of these events, with the current lack of radio detections strongly constraining the number of red giant companions in the progenitor systems (Chomiuk et al. 2016). In both cases prompt ($\leq 24$ hr) response is critical — because the velocity of the pre-explosion mass loss is much less than that of the outgoing supernova shock, in just one day the SN blastwave will sweep up years' to decades' worth of pre-explosion ejecta. With the advent of new wide-field optical surveys such as the Zwicky Transient Facility (Bellm 2014), discoveries of such very young SNe will become more routine, and so the need for ALMA follow-up will only increase.

Rapid-turnaround ToO observations may also be desirable to search for electromagnetic (EM) counterparts to gravitational wave events from Advanced LIGO/Virgo. In particular, binary neutron star mergers are expected to produce short $\gamma$-ray bursts (sGRBs), which will produce millimeter emission with a time delay determined by the observer's viewing angle (Nakar & Piran 2011; Metzger & Berger 2012). Surveying a full localization region ($\geq 100$ deg$^2$ for the foreseeable future) is infeasible given ALMA's small field of view, but ALMA provides sufficient sensitivity to detect a radio counterpart *if a precise localization is available from the initial identification of another EM counterpart*, such as a sGRB or a kilonova (Metzger & Berger 2012). Based on cm observations of on-axis *Swift* sGRBs (Fong et al. 2015), emission in the ALMA band should peak within 1–2 days, so turnaround within 24 hours is desirable for these high priority, rare events.

For TDEs, observations in millimeter are optimal for detecting the radiation directly from newly formed (possibly relativistic) jets in the earliest stages. Low frequency radio emission will be self-absorbed at early times (Irwin et al. 2015), while the infrared and optical emission suffers serious contamination due to emission and/or extinction in the host galaxies. Using the few examples we currently have in hand as a guide (e.g., *Swift* J1644+57: Zauderer et al. 2011), we expect the mm emission to peak on a time scale of a week. This does not necessitate new requirements on response time but represents yet another case where prompt data access is essential for planning follow-up.

In the planetary sciences, rapid-turnaround ToO observations are required to investigate transient events such as global Martian dust storms, stratospheric disturbances on the gas and ice giants, or the eruption of Europa's water vapor plumes. Observations of comet infall in planetary atmospheres sheds light on the role of the chemical contribution of cometary material to larger bodies. The observations of such events, which are mostly unpredictable and short-lived, may be somewhat more challenging to schedule because they can be subject to stringent visibility constraints as well as the requirement for rapid turnaround.

## 3.2 Recommended Policy Changes

We recommend that ALMA devise a policy to allow rapid analysis for ToO observations. Our ideal approach would be to simply allow immediate PI access to the relevant data products after QA0 verification. We strongly encourage the investigation of alternative schemes if ALMA cannot support this policy. One possible approach is that after a ToO proposal requiring rapid data analysis is approved, an ALMA analyst with relevant expertise be pre-selected to provide rapid QA and preliminary analysis in the event of a ToO trigger. We note that the most rapid analysis ($\lesssim 12$–$24$ h) is required only in the earliest days after trigger — as events age, the requirement on analysis turnaround time loosens.

To maximize the science impact of ALMA's *unique* ability to provide high-sensitivity, rapid-turnaround observations in the mm/sub-mm bands, we recommend that ToO triggers be prioritized ahead of engineering and array development activities to the greatest extent possible. To aid proposal planning and TAC evaluations, we recommend that future editions of the Proposer's Guide provide quantitative information, based on historical ALMA operations data, about the fraction of the time that ToO triggers could not have been honored due to engineering and development work.

We further recommend that ALMA establish protocols to ensure efficient communication between PIs and observatory staff when ToOs are triggered. Good communication is essential to maximize the scientific return of ToO projects in the face of changeable weather, multi-observatory coordination, and rapidly changing information. For instance, a PI may obtain a measurement of a GRB host galaxy redshift and therefore wish to adjust the frequency setup of their project in order to capture mm-band absorption lines from the host ISM. Such an adjustment is technically a "major change request" which would result in an unacceptable (multi-day) delay under "business as usual" conditions.

Finally, we recommend that ALMA endeavor to streamline response to ToO triggers such that the characteristic response time may be reduced from 48 to 24 hours, recognizing that the requested response time can never be fully guaranteed. If certain circumstances are more conducive to rapid ToO turnaround than others (e.g., weekdays versus weekends), future editions of the Proposer's Guide should outline the relevant factors and quantify both best-case and worst-case turnaround times. If there are concerns about very rapid ToO turnaround times leading to significant disruptions to standard observatory operations, we recommend that ALMA investigate a tiered scheme similar to that used by *Hubble*, where ToO requests with more stringent turnaround time requirements are classified as "disruptive," and only a limited number of such proposals are accepted each cycle.

## 4 Long, Continuous Observations

The C4PG states that "proposals that require Band 8 or better weather conditions for more than two hours continuously will be rejected on technical grounds. Observations with less stringent weather requirements are limited to three hours of continuous monitoring. The longest continuous observations allowed are 3 hours for Bands 3–7 and 2 hours for Bands 8–10." These limitations are substantially shorter than the longest possible track on the vast majority of potential ALMA targets.

### 4.1 Science Drivers

Long continuous observations are needed to study phenomena that are intrinsically variable on ~hour timescales. If such variability comes in the form of infrequent flares or other discrete events, long tracks are needed to provide context for any events that are seen. When ground- and space-based observatories perform coordinated observations of targets, long tracks are essential to take full advantage of the limited and valuable space observatory time.

As described above, last year a special DDT was granted for multiple, 6-hour continuous observations of Sgr A* — demonstrating the scientific value of this kind of observing mode. ALMA observations, combined with data from other observatories, could probe the physical origin of flares from Sgr A*. With only 2–4 flares per day, however, long tracks are required to yield a statistical sample of events with adequate multi-wavelength coverage. In particular, the 6-hour tracks obtained last year were useful, but the discontinuous coverage had one of the sub-tracks starting just at the beginning of a major X-ray+NIR event. This left the sub-mm state prior to the event undetermined, limiting the scientific conclusions despite the long overall observing duration relative to current ALMA policy.

Low-mass stars and sub-stellar objects can have rotational periods of ~hours, leading to comparable observational requirements. ALMA observations of ultra-cool dwarfs (objects having effective temperatures $\lesssim 2700$ K) have revealed that these objects can accelerate electrons to MeV energies (Williams et al. 2015), with potentially significant implications for the habitability of any planets orbiting such objects (such as those of the TRAPPIST-1 system; Gillon et al. 2017). Monitoring over multiple consecutive rotations — i.e., time scales of $\gtrsim 10$ hr — is necessary to disentangle variations due to the structure of the magnetosphere and those due to stochastic flares. A wide variety of solar system objects, ranging from asteroids such as Ceres to comets and the giant planets, have comparable rotation periods, and need near-continuous observations to derive longitudinal mapping of atmospheric and surface features.

## 4.2 Recommended Policy Changes

We recommend that the limitations on long continuous observations be lifted. Instead, ALMA should establish policies regarding when long observations may be interrupted, as well as how such observations are "retried" if they cannot be completed as initially planned. We suggest a simple prescription, such as one in which these observations are retried if less than 75% of the observation was completed and there are not external constraints that prevent a retry. However, a policy in which long observations are allowed but are simply abandoned if they cannot be completed (mirroring that used by many optical observatories) would be preferable to the current policy. Once again, if there are concerns that incomplete observations would lead to a decrease in ALMA's overall observing efficiency, we recommend that a fixed number of slots be reserved for such observations each cycle.

We recommend that the Observing Tool be modified to aid the planning of long, continuous observations by allowing users to optionally override the limitations on observing block lengths, even if such observations end up being implemented as multiple sequential observing blocks in practice. It is especially important that the tool properly estimate the reduced calibration overheads that are achievable in long, continuous observations.

# 5 Conclusion

We have presented three areas in which improved ALMA capabilities could dramatically increase its ability to produce unique, cutting-edge science. Although these areas are related by their connection to what we term "time-domain science," the astrophysical questions motivating our recommendations are broad, running the gamut from solar system science, to stellar astrophysics, to the study of cataclysms at the edge of the observable universe.

Our recommendations are efficient in the sense that they enable a significant increase in ALMA's science impact at *very low cost*, because they can be implemented without requiring the development of significant new technical capabilities. We therefore urge ALMA to prioritize their implementation.

# ATSIG Membership

| | |
|---|---|
| Kate D. Alexander | Harvard-Smithsonian Center for Astrophysics |
| Edo Berger | Harvard-Smithsonian Center for Astrophysics |
| Geoff Bower | Academica Sinica Institute of Astronomy and Astrophysics |
| Sarah Casewell | University of Leicester |
| S. Brad Cenko | Goddard Space Flight Center, NASA |
| Shami Chatterjee | Cornell University |
| Ilse Cleeves | Harvard-Smithsonian Center for Astrophysics |
| Jim Cordes | Cornell University |
| Jeremy Drake | Harvard-Smithsonian Center for Astrophysics |
| Maria Drout | Carnegie Observatories |
| Trent Dupuy | University of Texas at Austin |
| Tarraneh Eftekhari | Harvard-Smithsonian Center for Astrophysics |
| Giovanni Fazio | Harvard-Smithsonian Center for Astrophysics |
| Wen-fai Fong | University of Arizona |
| James Guillochon | Harvard-Smithsonian Center for Astrophysics |
| Mark Gurwell | Harvard-Smithsonian Center for Astrophysics |
| Michael Johnson | Harvard-Smithsonian Center for Astrophysics |
| Tomasz Kaminski | Harvard-Smithsonian Center for Astrophysics |
| Albert Kong | National Tsing Hua University |
| Tanmoy Laskar | National Radio Astronomy Observatory |
| Casey Law | University of California, Berkeley |
| Stuart P. Littlefair | University of Sheffield |
| Meredith MacGregor | Harvard-Smithsonian Center for Astrophysics |
| W. Peter Maksym | Harvard-Smithsonian Center for Astrophysics |
| Lynn Matthews | Harvard-Smithsonian Center for Astrophysics |
| Michael McCollough | Harvard-Smithsonian Center for Astrophysics |
| Stefanie Milam | Goddard Space Flight Center, NASA |
| Arielle Moullet | National Radio Astronomy Observatory |
| Matt Nicholl | Harvard-Smithsonian Center for Astrophysics |
| Aaron Rizzuto | University of Texas at Austin |
| Barry Rothberg | Large Binocular Telescope Observatory |
| Andrew Seymour | Arecibo Observatory |
| Eric Villard | Joint ALMA Observatory |
| Belinda Wilkes | Harvard-Smithsonian Center for Astrophysics |
| Peter K. G. Williams | Harvard-Smithsonian Center for Astrophysics |
| Steven Willner | Harvard-Smithsonian Center for Astrophysics |
| Farhad Yusuf-Zadeh | Northwestern University |